\documentclass[english]{article}
\usepackage[T1]{fontenc}
\usepackage[latin1]{inputenc}
\usepackage{graphicx}
\usepackage{amssymb}

\makeatletter

\usepackage{babel}
\makeatother
\begin{document}

\title{Integral representation of one dimensional three particle scattering
for $\delta$ function interactions}

\author{A. Amaya-Tapia$^{1}$, G. Gasaneo$^{2,3}$, S. Ovchinnikov$^{4}$,
J. H. Macek$^{2,4}$ and S. Y. Larsen$^{5}$}

\maketitle
\centerline{$^{1}$ Centro de Ciencias F\'{\i}sicas, UNAM, AP 48-3,Cuernavaca,
Mor. 62251, M\'{e}xico.} \centerline{$^{2}$ Department of Physics
and Astronomy, University of Tennessee, Knoxville, TN 37996-1501,
USA.} \centerline{$^{3}$ Departamento de F\'{\i}sica, Universidad
Nacional del Sur, Av. Alem 1253, (8000) Bah\'{\i}a Blanca, Buenos
Aires, Argentina.} \centerline{$^{4}$ Oak Ridge National Laboratory,
PO Box 2008, Oak Ridge, Tennessee 37831, USA} \centerline{$^{5}$
100 Forest Place, Apt. 1305, Oak Park, IL 60301, USA.}

\begin{abstract}
The Schr\"{o}dinger equation, in hyperspherical coordinates, is solved
in closed form for a system of three particles on a line, interacting
via pair delta functions. This is for the case of equal masses and
potential strengths. The interactions are replaced by appropriate
boundary conditions. This leads then to requiring the solution of
a free-particle Schr\"{o}dinger equation subject to these boundary
conditions. A generalized Kontorovich - Lebedev transformation is
used to write this solution as an integral involving a product of
Bessel functions and pseudo-Sturmian functions. The coefficient of
the product is obtained from a three-term recurrence relation, derived
from the boundary condition. The contours of the Kontorovich-Lebedev
representation are fixed by the asymptotic conditions. The scattering
matrix is then derived from the exact solution of the recurrence relation.
The wavefunctions that are obtained are shown to be equivalent to
those derived by McGuire. The method can clearly be applied to a larger
number of particles and hopefully might be useful for unequal masses
and potentials. 
\end{abstract}

\section*{Introduction}

Three-body systems and processes are of fundamental interest in physics
\cite{Evora}. One of these, with which a number of us have been concerned,
is the recombination of three-particles to a dimer plus a free particle,
in a many body system forming a Bose-Einstein condensate \cite{NielsenMacek}.
The condensate is not the lowest state of the system, but a metastable
state. The 3-body recombination is the dominant mechanism for cooling
and lowering the overall energy of the system.

Experimental and theoretical studies have shown that this recombination
rate depends mainly on the two-body scattering length $a$ \cite{Inouye,Moerdijk,Esry,Fedichev,NielsenMacek},
as the collision energy is low and the interaction is weak - owing
to large interparticle distances, and on the bound state energies.

This would suggest that zero-range potentials (ZRP), defined in terms
of the scattering length \cite{DemkovOstrovskii},

\begin{equation}
\lim_{r\rightarrow0}\left[\frac{1}{r\psi}\frac{\partial\left(r\psi\right)}{\partial r}\right]=-1/a.\label{i0.1}\end{equation}
 can be applied to model the interaction between the particles of
the condensate. It has been shown by Nielsen and Macek using the hidden
crossing technique that the ZRP describes properly the recombination
transition in a system of three $^{4}$He atoms \cite{NielsenMacek}.
Also, Gasaneo and Macek showed that the ZRP gives a quite good representation
for the adiabatic potential of the same system \cite{GasaneoMacek}.
A closed form solution for a system of three-particles interacting
via a ZRP has been recently presented by Gasaneo et al \cite{GasaneoPRA}.
The fragmentation process \ $^{4}$He$_{2}+~^{4}$He$\longrightarrow~^{4}$He$+~^{4}$He$+~^{4}$He
was studied and relatively good agreement was found when compared
with the hidden crossing calculations.

In this paper, we seek to apply our techniques to a famous model:
3 particles in one dimension, subject to pair delta-function interactions.
For this model, introduced by McGuire \cite{McG}, one can obtain
exact solutions for the wave functions, the scattering matrix and
the binding energies, in the case of particles of identical masses
and equally weighted interactions. As such it has been extended to
a larger number of particles \cite{Yang}, using Bethe's Ansatz \cite{Bethe},
and also found to be exceedingly useful when used as a test-bed for
the development of a number of different methods (pertubative, Faddeev,
hyperspherical adiabatic, etc.) \cite{numerous}.

Here, we note that using ZRP and (\ref{i0.1}), in 3-dimensions, leads
to the Thomas effect and the collapse of the 3-body ground state \cite{thomas}.
However, in one dimension, an equation similar to (\ref{i0.1}) -
with $a$ not the scattering length - provides boundary conditions
which correctly characterizes the wave functions and replace the use
of the $\delta$-function interactions, and should therefore again
give us exact results. One of these, though, is that the recombination
rate, for this model, is exactly zero.

In section II we propose a solution, written in integral form, for
the free particle Schr\"{o}dinger equation, written in hyperspherical
coordinates. A linear combination of free particle solutions can then
be found to satisfy the boundary condition that we alluded to earlier,
and thus provide us with the solution of the problem with interaction.
The requirement that the wave function satisfy the boundary conditions
leads us to one of the important results of this paper, namely that
the weight of the free particle solutions, in the integral form, satisfies
a recurrence relation similar to that obtained in the refs. \cite{GOM}
and \cite{GasaneoPRA}.

In section III the method is applied to a particular case in which
two of the particles are bound. It is shown that the recurrence relation,
defining the coefficient of the free-particle expansion, can be solved
in closed form and, thus, the scattering matrix is also obtained in
a closed form. This allows us to have a detailed test of our method.
In this section it is also shown that the wave function obtained is
equivalent to the McGuire plane wave solution, and that our expression
for the ${\mathcal{S}}$ matrix is the matrix obtained by McGuire,
in the particular case discussed in this paper. In section IV the
relation between the hyperspherical adiabatic approach and the present
one is discussed.

In Appendix A, the pseudo-Sturmian functions are derived. In Appendix
B, the wave function is written as the symmetric wave plane in cartesian
coordinates.

\section*{Exact Integral Representation}

To begin the study of the three identical-particle system (therefore
with equal masses), consider the center of mass and Jacobi coordinates,
\begin{eqnarray}
r & = & \frac{1}{3}(x_{1}+x_{2}+x_{3}),\nonumber \\
\eta & = & \sqrt{\frac{1}{2}}(x_{1}-x_{2}),\nonumber \\
\xi & = & \sqrt{\frac{2}{3}}\left(\frac{x_{1}+x_{2}}{2}-x_{3}\right)\end{eqnarray}

\begin{figure}

\caption{One of three sets of Jacobi coordinates for the three particles.}
\includegraphics[width=7cm]{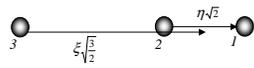}

\label{Fig1}
\end{figure}
 the $x_{i}\,$ give us the locations of the 3 particles along the
line, see Fig. 1 . Using polar coordinates, the 2 Jacobi variables
allow us to define, in turn, a hyper radius $R$ and an angle $\theta$
as \begin{equation}
\eta=R\cos\theta\qquad\qquad\xi=R\sin\theta\label{co}\end{equation}
 where $-\pi<\theta\leq\pi\,\,$and\thinspace $0\leq R<\infty$.
In terms of these coordinates the Schr\"{o}dinger equation for the
`relative' system can be written as

\begin{equation}
H\Psi\left(R,\theta\right)=(\frac{2m}{\hbar^{2}})\, E\Psi\left(R,\theta\right)\label{sr}\end{equation}
 where

\begin{equation}
H=-\left(\frac{1}{R}\frac{\partial}{\partial R}R\frac{\partial}{\partial R}+\frac{1}{R^{2}}\frac{\partial^{2}}{\partial\theta^{2}}\right)+\frac{1}{R}C\left(\theta\right).\end{equation}
 The function $C\left(\theta\right)$ is defined by \begin{equation}
C\left(\theta\right)=\frac{\pi}{3}c\sum\limits _{j=0}^{5}\,\delta\left(\theta-\theta_{j}\right),\end{equation}
 where the coefficient $c$ equals $(3/\pi\sqrt{2})(2m/\hbar^{2})g$,
$g$ being the strength of the interactions. This $c$ is negative
for attractive interactions and positive for repulsive ones. The angles
$\theta_{j}$ equal $(2j+1)\pi/6$. The lines $\theta=\theta_{j}$
divide the $(\rho,\theta)$ plane in six regions. In each region the
order of particles is fixed, so that between $\theta_{4}$ and $\theta_{5}$,
$x_{1}<x_{2}<x_{3}$, etc. A different permutation of particles is
associated to each region. >From now on, we will choose the units
such that $2m=1$ and $\hbar=1$. In each sector, we now seek a free
particle solution that satisfies the boundary condition that will
replace the effect of the potential, i.e. 

\begin{equation}
\lim_{\theta^{-}\rightarrow\theta_{j}}\left[\frac{1}{R\Psi(R,\theta)}\frac{\partial\Psi(R,\theta)}{\partial\theta}\right]=-\frac{1}{a},\label{wfbc}\end{equation}
 where $\theta^{-}=\theta<\theta_{j}\; j=0,1,...,5$. In Eq.(\ref{wfbc})
$a=(6/\pi c)$ does not depend on $j$ because all the strengths of
the interactions and all the masses are equal. Writing a solution
for the free particle system as the product $\psi_{free}\left(R,\theta\right)=\Theta\left(\nu,\theta\right)R^{1/2}{\mathcal{R}}_{\nu}\left(\mathrm{K}R\right)$,
where $\mathrm{K}^{2}=E$, leads to the set of free particle equations

\begin{eqnarray}
 &  & \left.\frac{R^{2}}{R^{-1/2}Z_{\nu}\left(\mathrm{K}R\right)}\left[\frac{\partial^{2}}{}{\partial R^{2}}+\mathrm{K}^{2}\right]R^{-1/2}Z_{\nu}\left(\mathrm{K}R\right)\right.\nonumber \\
 &  & \left.=\frac{-1}{\Theta(\nu,\theta)}\left(\frac{\partial^{2}}{}{\partial\theta^{2}}+\frac{1}{4}\right)\Theta(\nu,\theta)=\nu^{2}-\frac{1}{4}\right.,\end{eqnarray}
 where $Z_{\nu}\left(\mathrm{K}R\right)=R^{1/2}{\mathcal{R}}_{\nu}\left(\mathrm{K}R\right)$
is a Bessel function and $\nu$ a separation constant. If for the
$\Theta\left(\nu,\theta\right)$ functions we choose the pseudo-Sturmian
functions $S(\nu,\theta)$, defined, for fixed $\nu$, as the solutions
of \begin{equation}
-\left[\frac{\partial^{2}}{\partial\theta^{2}}+\frac{1}{4}-\rho\left(\nu\right)C\left(\theta\right)\right]S(\nu,\theta)=\left(\nu^{2}-\frac{1}{4}\right)S(\nu,\theta),\label{seq}\end{equation}
 then the functions $\psi_{free}\left(\theta,R\right)=S(\nu,\theta)Z_{\nu}\left(\mathrm{K}R\right)$
are solutions of the Schr\"{o}dinger equation Eq.(\ref{sr}) for
values of $\rho\left(\nu\right)=R.$ Note that Eq.(\ref{seq}) may
be replaced by the relation

\begin{equation}
\left[\frac{\partial^{2}}{\partial\theta^{2}}+\nu^{2}\right]S(\nu,\theta)=0\label{afree}\end{equation}
 subject to the boundary conditions

\begin{equation}
\lim_{\theta^{-}\rightarrow\theta_{j}}\left[\frac{1}{\rho\left(\nu\right)}\frac{1}{S(\nu,\theta)}\frac{\partial S}{\partial\theta}(\nu,\theta)\right]=-\frac{1}{a},\; j=0,1,...,5.\label{sbc}\end{equation}
 In the last equation, we assumed that $S(\nu,\theta)$ is symmetric
about each line $\theta=\theta_{j}$.

We now propose to write the general wave function of the system as
a Kontorovich-Lebedev transform, in terms of the base functions just
discussed, that is, as

\begin{equation}
\Psi(R,\theta)=\int_{\varsigma}d\nu A(\nu)S\left(\nu,\theta\right)Z_{\nu}(\mathrm{K}R),\label{klr}\end{equation}
 provided that its derivative satisfies the boundary conditions, i.e.
Eq.(\ref{wfbc}). The contour of integration must be chosen so that
the wave function has the correct asymptotic behaviour.

Following the reasoning of Gasaneo et al \cite{GasaneoMacek}, we
will now show that the boundary conditions, Eq.(\ref{wfbc}), $\,\,$can
be transformed into a recurrence relation for $A\left(\nu\right)$.
First, we substitute Eq.(\ref{klr}) in Eq.(\ref{wfbc}), and then
interchange the order in which the integral and the derivative are
taken, to obtain for each $j$\begin{eqnarray}
 &  & \left.\lim_{\theta^{-}\rightarrow\theta_{j}}\int_{\varsigma}d\nu A(\nu)\left[\frac{1}{R}Z_{\nu}(\mathrm{K}R)\frac{\partial S(\nu,\theta)}{\partial\theta}\right.\right.\nonumber \\
 &  & \left.\left.+\frac{1}{a}S(\nu,\theta)Z_{\nu}(\mathrm{K}R)\right]=0.\right.\end{eqnarray}
 Second, we use Eq.(\ref{sbc}) and the identity $(2\nu/z)Z_{\nu}(z)$
$=Z_{\nu+1}(z)-Z_{\nu-1}(z)$ to transform the equation to \begin{eqnarray}
\lim_{\theta^{-}\rightarrow\theta_{j}} &  & \left\{ \int_{\varsigma}d\nu A(\nu)\frac{1}{\nu}\left[-\rho\left(\nu\right)/a\right]S(\nu,\theta)\right.\nonumber \\
 &  & \times\left[Z_{\nu+1}(KR)-Z_{\nu-1}(KR)\right]\nonumber \\
 &  & \left.\left.+\frac{2}{Ka}\int_{\varsigma}d\nu A(\nu)S(\nu,\theta)Z_{\nu}(KR)\right\} =0.\right.\end{eqnarray}
 We assumed in the previous equation that $\mathrm{K}=iK$ and $K\geq0$,
because we are mainly interested in negative energies. By selecting
the appropriate contours we can now transform the last equation to

\begin{eqnarray}
\lim_{\theta^{-}\rightarrow\theta_{j}} &  & \int_{\varsigma}d\nu\left[A(\nu-1)\frac{1}{\nu-1}\rho(\nu-1)S(\nu-1,\theta)\right.\nonumber \\
 &  & -A(\nu+1)\frac{1}{\nu+1}\rho(\nu+1)S(\nu+1,\theta)\nonumber \\
 &  & \left.\left.-\frac{2}{K}A(\nu)S(\nu,\theta)\right]Z_{\nu}(KR)=0\right.\end{eqnarray}

Since the set of Bessel functions forms a complete set of basis functions,
the function within the square brackets should be zero at the limit.
We arrive, finally, at the recurrence relation that we are looking
for,

\begin{eqnarray}
 &  & B(\nu-1)\rho(\nu-1)S(\nu-1,\theta_{j})\label{rr}\\
 &  & \left.-B(\nu+1)\rho(\nu+1)S(\nu+1,\theta_{j})=\frac{2\nu}{K}B(\nu)S(\nu,\theta_{j})\right.\nonumber \end{eqnarray}
 where $B(\nu)=A(\nu)/\nu$. In the following section, we will apply
this approach to a particular case of this three body system and show
that we can obtain the wave function and the $\mathcal{S}$-matrix.

\section*{2+1 System}

Consider now the case where two of the particles are bound. The wave
function $\psi(R,\theta)$ can still be written in terms of the Kontorovich-Lebedev
representation, Eq.(\ref{klr}). The unnormalized angle pseudo-Sturmian
function $S(\nu,\theta)$, a six-fold symmetric function, is defined
by the Eqs.(\ref{afree}) and (\ref{sbc}). As can be seen in Appendix
A, the function $S(\nu,\theta)$ may written as

\begin{equation}
S(\nu,\theta)=\cos\left[\left(\theta-j\frac{\pi}{3}\right)\nu\right]\qquad\left|\theta-j\frac{\pi}{3}\right|<\frac{\pi}{6},\;\label{st}\end{equation}
 with $j=0,1,...,5$, where $\rho(\nu)$ satisfies the relation

\begin{equation}
\nu\tan(\nu\frac{\pi}{6})=\frac{1}{\left(6/\pi c\right)}\,\rho(\nu).\label{rnu}\end{equation}

>From the previous section, we can immediately conclude that $A(\nu)$
satisfies the recurrence relation

\begin{eqnarray}
 &  & A(\nu+1)\sin\left[\left(\nu+1\right)\frac{\pi}{6}\right]-A(\nu-1)\sin\left[\left(\nu-1\right)\frac{\pi}{6}\right]\nonumber \\
 &  & =-\frac{\pi c}{3K}A(\nu)\cos\left[\nu\frac{\pi}{6}\right],\label{rec1}\end{eqnarray}
.

\subsection*{Solution of the recurrence relation}

The recurrence relation, displayed in Eq.(\ref{rec1}), can be written
as \begin{eqnarray}
 &  & e^{i\left(\pi/6\right)\nu}\left[A\left(\nu+1\right)e^{i\left(\pi/6\right)}-A(\nu-1)e^{-i\left(\pi/6\right)}\right.\nonumber \\
 &  & \left.+\frac{i\pi c}{3K}A(\nu)\right]+e^{-i\left(\pi/6\right)\nu}\left[-A\left(\nu+1\right)e^{-i\left(\pi/6\right)}\right.\nonumber \\
 &  & \left.\left.+A(\nu-1)e^{i\left(\pi/6\right)}+\frac{i\pi c}{3K}A(\nu)\right]=0.\right.\label{rec2}\end{eqnarray}
 An inspection, of the solution of the recurrence relation - Eq.(25)
in ref. \cite{GasaneoMacek}, leads us to propose a coefficient in
the form of the series

\begin{eqnarray}
A\left(\nu\right) & = & e^{-\beta\nu}\left[e^{-i\left(\pi/3\right)\nu}+\mathcal{S}e^{i\left(\pi/3\right)\nu}\right.\nonumber \\
 &  & \left.+\mathcal{S}_{1}e^{-i\left(\pi/6\right)\nu}+\mathcal{S}_{2}e^{i\left(\pi/6\right)\nu}+\mathcal{S}_{3}\right]\label{anu}\end{eqnarray}
 Substituting this expression in Eq.(\ref{rec2}), and equating to
zero the coefficients of exponentials, with different arguments that
depend on $\nu$, we obtain the following values for the parameters:

\begin{eqnarray}
\mathcal{S} & = & \tan\left(\frac{\pi}{6}-i\beta\right)\cot\left(\frac{\pi}{6}+i\beta\right)\nonumber \\
\mathcal{S}_{3} & = & -\cot\frac{\pi}{6}\cot\left(\frac{\pi}{6}+i\beta\right)\label{par}\end{eqnarray}
\begin{eqnarray}
\cos\left(i\beta\right) & = & -\frac{\pi c}{6K}\nonumber \\
\sin\left(i\beta\right) & = & i\,\frac{k}{K}.\label{beta}\end{eqnarray}
 Consequently, the solution for the coefficient can be written as

\begin{equation}
A\left(\nu\right)=e^{-\beta\nu}\left(e^{\left(-i\pi/3\right)\nu}+\mathcal{S}\, e^{\left(i\pi/3\right)\nu}+\mathcal{S}_{3}\,\right),\label{a}\end{equation}
 or

\begin{equation}
A\left(\nu\right)=2\, e^{-\beta\nu}\left[\cos\left(\frac{\pi}{3}\nu+\delta\right)+\alpha\right]\end{equation}
 where

\begin{equation}
\mathcal{S}=e^{2i\delta}\end{equation}
 and

\begin{equation}
\alpha=-\frac{1}{2}\cot\frac{\pi}{6}\sqrt{\cot\left(\frac{\pi}{6}-i\beta\right)\cot\left(\frac{\pi}{6}+i\beta\right)}\end{equation}

In the next section, we demonstrate that $\mathcal{S}$ represents
the scattering matrix and, accordingly, $\delta$ the phase shift.
We should stress the remarkable fact that the S-matrix appears explicitly
in the solution of the recurrence relation. In the next subsection
it is shown that the expression obtained for $\mathcal{S}$ in this
work is equivalent to the formula for the exact symmetric $\mathcal{S}$-matrix
for the 2 + 1 process, given in Ref. \cite{Amaya}.

\subsection*{Asymptotic wave function}

To be specific we will restrict the following discussion to the case
of total negative energies, and will write $K=\sqrt{(\pi c)^{2}/36-k^{2}}\geq0$,
in which $-(\pi c)^{2}/36$ is the two-body bound energy and $k^{2}$
is the effective energy. Next, we will show that the imaginary axis
is the appropriate contour to obtain the correct asymptotic behaviour
of the wave function. Substituting the coefficients $A(\nu)$ defined
in Eqs.(\ref{par}) and (\ref{a}), the pseudo-Sturmian functions
given in Eq.(\ref{st}) and the modified Bessel functions $K_{\nu}(KR)$,
into Eq.(\ref{klr}), as well as choosing the imaginary axis as the
contour of integration, we find \begin{eqnarray}
 &  & \Psi=\nonumber \\
 &  & \int_{\varsigma}d\nu\left(\cosh\left[\left(i\pi/3+\beta\right)\nu\right]-\sinh\left[\left(i\pi/3+\beta\right)\nu\right]\right)\nonumber \\
 &  & \times\cos\left[\left(\theta-j\frac{\pi}{3}\right)\nu\right]K_{\nu}(KR)\nonumber \\
 &  & +\mathcal{S}\int_{\varsigma}d\nu\left(\cosh\left[\left(i\pi/3-\beta\right)\nu\right]+\sinh\left[\left(i\pi/3-\beta\right)\nu\right]\right)\nonumber \\
 &  & \times\cos\left[\left(\theta-j\frac{\pi}{3}\right)\nu\right]K_{\nu}(KR).\nonumber \\
 &  & +\mathcal{S}_{3}\int_{\varsigma}d\nu\left(\cosh\left[\beta\nu\right]-\sinh\left[\beta\nu\right]\right)\nonumber \\
 &  & \times\cos\left[\left(\theta-j\frac{\pi}{3}\right)\nu\right]K_{\nu}(KR).\end{eqnarray}
 Note that in the above expression the exponentials in the coefficients
have been written in terms of hyperbolic functions. The integral over
the odd terms vanishes, leaving only the even terms in the integrand.
After a trigonometric identity, this leads to

\begin{eqnarray}
 &  & \Psi=\nonumber \\
 &  & \frac{1}{2}\left\{ \int_{\varsigma}d\nu\cosh\left(\left[\beta+i\left(\theta-\left[j-1\right]\frac{\pi}{3}\right)\right]\nu\right)K_{\nu}(KR)\right.\nonumber \\
 &  & +\int_{\varsigma}d\nu\cosh\left(\left[-\beta+i\left(\theta-\left[j+1\right]\frac{\pi}{3}\right)\right]\nu\right)K_{\nu}(KR)\nonumber \\
 &  & +\mathcal{S}\left[\int_{\varsigma}d\nu\cosh\left(\left[-\beta+i\left(\theta-\left[j-1\right]\frac{\pi}{3}\right)\right]\nu\right)K_{\nu}(KR)\right.\nonumber \\
 &  & +\;\left.\int_{\varsigma}d\nu\cosh\left(\left[\beta+i\left(\theta-\left[j+1\right]\frac{\pi}{3}\right)\right]\nu\right)K_{\nu}(KR)\right]\nonumber \\
 &  & +\mathcal{S}_{3}\left[\int_{\varsigma}d\nu\cosh\left(\left[-\beta+i\left(\theta-j\frac{\pi}{3}\right)\right]\nu\right)K_{\nu}(KR)\right.\nonumber \\
 &  & +\left.\int_{\varsigma}d\nu\cosh\left(\left[\beta+i\left(\theta-j\frac{\pi}{3}\right)\right]\nu\right)K_{\nu}(KR)\right]\end{eqnarray}
 Using the Kontorovich-Levedev Transforms \cite{Gradshteyn}, we obtain

\begin{eqnarray}
\Psi & = & \frac{i\pi}{2}\left(\exp\left\{ -KR\cosh\left(\beta+i\left[\theta-\left(j-1\right)\frac{\pi}{3}\right]\right)\right\} \right.\nonumber \\
 &  & +\exp\left\{ -KR\cosh\left(-\beta+i\left[\theta-\left(j+1\right)\frac{\pi}{3}\right]\right)\right\} \nonumber \\
 &  & +\mathcal{S\,}\left[\exp\left\{ -KR\cosh\left(\beta+i\left[\theta-\left(j+1\right)\frac{\pi}{3}\right]\right)\right\} \right.\nonumber \\
 &  & \left.+\exp\left\{ -KR\cosh\left(-\beta+i\left[\theta-\left(j-1\right)\frac{\pi}{3}\right]\right)\right\} \right]\nonumber \\
 &  & +\mathcal{S}_{3}\left[\exp\left\{ -KR\cosh\left(-\beta+i\left[\theta-j\frac{\pi}{3}\right]\right)\right\} \right.\nonumber \\
 &  & \left.+\exp\left\{ -KR\cosh\left(\beta+i\left[\theta-j\frac{\pi}{3}\right]\right)\right\} \right]\end{eqnarray}

Introducing $\beta$ from Eq.(\ref{beta}) into this expression, yields

\begin{center}\begin{eqnarray}
\Psi & =\frac{i\pi}{2} & \left(\exp\left\{ \frac{\pi c}{6}R\cos\left[\theta-\left(j-1\right)\frac{\pi}{3}\right]\right.\right.\nonumber \\
 &  & \left.-ikR\sin\left[\theta-\left(j-1\right)\frac{\pi}{3}\right]\right\} \nonumber \\
 &  & +\exp\left\{ \frac{\pi c}{6}R\cos\left[\theta-\left(j+1\right)\frac{\pi}{3}\right]\right.\nonumber \\
 &  & \left.ikR\sin\left[\theta-\left(j+1\right)\frac{\pi}{3}\right]\right\} \nonumber \\
 & +\mathcal{S\,} & \left[\exp\left\{ \frac{\pi c}{6}R\cos\left[\theta-\left(j+1\right)\frac{\pi}{3}\right]\right.\right.\nonumber \\
 &  & \left.-ikR\sin\left[\theta-\left(j+1\right)\frac{\pi}{3}\right]\right\} \nonumber \\
 &  & +\exp\left\{ \frac{\pi c}{6}R\cos\left[\theta-\left(j-1\right)\frac{\pi}{3}\right]\right.\nonumber \\
 &  & \left.\left.ikR\sin\left[\theta-\left(j-1\right)\frac{\pi}{3}\right]\right\} \right]\nonumber \\
 & +\mathcal{S}_{3} & \left[\exp\left\{ \frac{\pi c}{6}R\cos\left[\theta-j\frac{\pi}{3}\right]\right.\right.\nonumber \\
 &  & \left.ikR\sin\left[\theta-j\frac{\pi}{3}\right]\right\} \nonumber \\
 &  & +\exp\left\{ \frac{\pi c}{6}R\cos\left[\theta-j\frac{\pi}{3}\right]\right.\nonumber \\
 &  & \left.\left.\left.-ikR\sin\left[\theta-j\frac{\pi}{3}\right]\right\} \right]\right),\nonumber \\
 &  & j=0,1,...,5\label{symf}\end{eqnarray}
\end{center}

From Eq.(\ref{st}) it is easily seen that this wave function is fully
symmetric under the interchange of particles. The function is invariant
under the addition of $\pi/3$ to $\theta$, together with the addition
of one unit to $j,$ which is what should be done to move from one
region in the $\left(\rho,\theta\right)$plane to its next counterclockwise
neighbour. Remember that for each region there is a specific order
of the particles.

We can also see that the real part of each of the exponential arguments
is negative, except when $\theta-\left(j\mp1\right)\pi/3=\pm\pi/2,$
that is on the lines $\theta=\theta_{j},$ where its value is zero.
Thus, when $R$ is large, the wave function is negligible except near
the lines $\theta=\theta_{j}.$ Note that only the first 4 terms give
a significant contribution in the asymptotic region. We can be conclude
that the form of the wave function is that of products of bound state
functions, associated with two particles, with oscillatory functions,
which describe the location of the third particle with respect to
the 2 bound ones. Evaluating, then, the wave function for large values
of $R$, its asymptotic form can be written as \[
\Psi(R,\theta')\sim e^{\left\{ \frac{\pi c}{6}R\cos\theta'\right\} }\left(e^{\left\{ -ikR\sin\theta'\right\} }+\mathcal{S\,}e^{\left\{ ikR\sin\theta'\right\} }\right)\]
 where $\pi/6<\theta'=\theta-\left(j-1\right)\frac{\pi}{3}<\pi/2,\; j=0,1,...,5$.
This asymptotic expression consists of a wave representing a two particles
bound state multiplied by an incoming wave, together with an outgoing
wave multiplied by ${\mathcal{S}}$.

>From the expression in Eq.(\ref{par}) the matrix $\mathcal{S}$
can be written as

\begin{eqnarray}
S & = & \frac{\sin\left(\frac{\pi}{6}+i\beta\right)\cos\left(\frac{\pi}{6}-i\beta\right)}{\cos\left(\frac{\pi}{6}+i\beta\right)\sin\left(\frac{\pi}{6}-i\beta\right)}\nonumber \\
 & = & \frac{\sin\frac{\pi}{3}-\sin\left(-2i\beta\right)}{\sin\frac{\pi}{3}+\sin\left(-2i\beta\right)}\nonumber \\
 & = & \frac{\sin\frac{\pi}{3}+2\sin\left(i\beta\right)\cos\left(i\beta\right)}{\sin\frac{\pi}{3}-2\sin\left(i\beta\right)\cos\left(i\beta\right)}.\end{eqnarray}

In terms of $K$,

\begin{equation}
\cos\left(i\beta\right)\sin\left(i\beta\right)=-\frac{\pi c}{6K}\,\frac{ik}{K}=\frac{-i\pi ck}{6\left(\pi^{2}c^{2}/36-k^{2}\right)}.\end{equation}
 Therefore \begin{equation}
\mathcal{S}=\frac{1-36\left(k/\pi c\right)^{2}-i\left(24/\sqrt{3}\right)\left(k/\pi c\right)}{1-36\left(k/\pi c\right)^{2}+i\left(24/\sqrt{3}\right)\left(k/\pi c\right)}\end{equation}
 This is, precisely, the scattering matrix

\begin{equation}
S=\frac{\left[-1-i\left(6\sqrt{3}/\pi c\right)k\right]\left[3+i\left(6\sqrt{3}/\pi c\right)k\right]}{\left[3+i\left(6\sqrt{3}/\pi c\right)k\right]\left[-1+i\left(6\sqrt{3}/\pi c\right)k\right]},\label{s}\end{equation}
 given as Eq.(61) in Ref. \cite{Amaya}. It corresponds to the symmetric
S matrix calculated for the specific process 2 +1.

The matrix $\mathcal{S}_{3}$, see again (\ref{par}), has the following
form as a function of k \begin{equation}
\mathcal{S}_{3}=\frac{3+i\left(6\sqrt{3}/\pi c\right)k}{}{-1+i\left(6\sqrt{3}/\pi c\right)k}.\end{equation}

$\mathcal{S}_{3}$ multiplies the shorter ranged part of the exact
wave function, that goes to zero when $R$ goes to $\infty$.

Additional insight can be gained, by following the reasoning of McGuire
\cite{McG}. The scattering of 3 asymptotically free particles, to
3 also asymptotically free particles, requires 3 (successive) collisions,
and yields the part of the wave function associated with the calculation
of the S-matrix. Intermediate stages, associated with fewer collisions,
give rise to the shorter ranged part of the wave functions. A similar
reasoning holds for the $2+1$ processes.

In conclusion, we have shown that this integration contour, and the
choice of Bessel functions, have imparted the correct asymptotic behaviour.
Furthermore, we can deduce from the asymptotic expression that the
coefficient $\mathcal{S}$ represents the S-matrix.

\section*{Relation to Adiabatic Theory}

The eigenfunctions of the following eigenvalue equation \cite{Amaya}
form a complete set of orthogonal hyperspherical adiabatic basis functions.
Changing, a bit, the usual notation:

\begin{equation}
\left[\frac{1}{R'^{2}}\left(\frac{\partial^{2}}{\partial\theta^{2}}+\frac{1}{4}\right)-\frac{1}{R'}C\left(\theta\right)+\Lambda_{\kappa}\left(R'\right)\right]B_{\kappa}(\theta;R')=0,\label{adi}\end{equation}
 where $R'$, a real parameter in this equation, is held fixed; $\kappa=0,6,12,...$
and $\Lambda_{\kappa}\left(R'\right)\rightarrow(\kappa^{2}-1/4)/R'^{2}$
as the interaction is turned off. The unnormalized eigenfunctions
\begin{equation}
B_{\kappa}(\theta;R')=\cos\left[q_{\kappa}\left(\theta-\frac{j\pi}{}{3}\right)\right],\end{equation}
 where $j$ is an integer such that $\left|\theta-j\frac{\pi}{}{3}\right|<\frac{\pi}{6}$
and $q_{\kappa}$ satisfies

\begin{equation}
q_{\kappa}\tan\left(\frac{\pi}{6}q_{\kappa}\right)=\frac{\pi R'c}{6}.\end{equation}
 In the adiabatic approach the parameter $R'$ is identified with
the hyper radius $R$. For $E<0$ there is only one open channel,
labeled by $\kappa=0$. For large $R$, the channel function is concentrated
along the lines defined by $\theta=\theta_{j}$. Accordingly it can
represent the two-body bound state. For $E>0$, there is an infinite
number of open channels, labeled by the successive numbers $\kappa$,
equal and greater than zero. They describe, asymptotically, three
free particles or a two-body bound state, together with a free particle.

Note that the function $\rho(\nu)$, Eq.(\ref{rnu}), is a real function
if, and only if, $\nu$ takes on values along the imaginary or the
real axis, see Fig. 2. It can be seen that the pseudo-Sturmian function
defined in Eq.(\ref{seq}) coincides, apart from normalization constants,
with the lowest adiabatic function $B_{0}(\theta;R')$ when $\nu=q_{0}$
is an imaginary number and $\rho(\nu)=R'(q_{0})$. Also, if $\nu=q_{\kappa}$
are in the real intervals $\left(3+\left[\kappa-6\right],9+\left[\kappa-6\right]\right)$
with $\kappa=6,12,...$, then $\rho(\nu)=R'(q_{\kappa})$ and the
pseudo-Sturmian functions become equal, except for the normalization
constants, to the adiabatic eigenfunctions $B_{\kappa}(\theta;R')$.

Thus, in the case of the example considered in this paper, that is
$E<0$ and the $2+1$ system, the integral Eq.(\ref{klr}), along
the imaginary axis in the complex $\nu$ plane, can be written in
terms of the lowest adiabatic function as

\begin{equation}
\Psi(R,\theta)=\int_{\varsigma}d\nu A(\nu)B_{0}\left(\theta;R'(\nu)\right)Z_{\nu}(KR),\label{intad}\end{equation}

\noindent where $\nu$ runs from $-i\infty$ to $i\infty$ . The most
important contribution of the adiabatic functions to the integral,
at large $R$, comes from the lines $\theta=\theta_{j}$, where two
of the particles are joined. When these adiabatic functions are multiplied
by the appropriate Bessels functions, their linear combination (Eq.
(\ref{intad})) should have the correct asymptotic behaviour, and
will represent a two-body bound system in the colliding with a third
particle.

\begin{figure}

\caption{Plot of the pseudo-Sturmian eigenvalue $\rho(\nu)$. In (a) we plot
$\rho(\nu)$ as a function of $\nu^{2}$. In (b) the plot of (a) is
rotated and flipped to give $\nu^{2}$ as a function of $\rho.$ For
$\rho$ positive, $\nu^{2}=\Lambda(\rho)\rho^{2}+\frac{1}{4}$.}
\includegraphics[width=7cm]{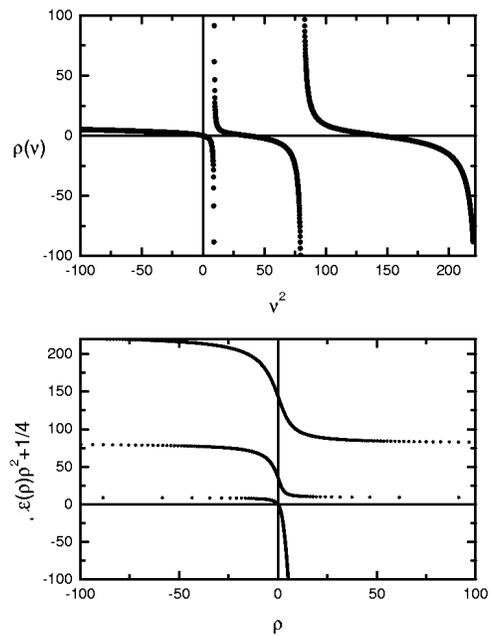}

\label{Fig2}
\end{figure}

\section*{Conclusions and Outlook}

We have shown that the integral representation approach within the
hyperspherical context, when applied to McGuire's model, offers a
reliable tool to study the collisional dynamics of the 3-body system.
We have obtained several interesting results, namely:

-An exact solution to the corresponding Schr\"{o}dinger equation.

-A closed form for the angular basis for this system, the pseudo-Sturmian
functions.

-A recurrence relation for the coefficients, in the expansion of the
wave function in terms of the free-particle basis.

-The S-matrix, obtained directly from the solution of the recurrence
relation.

-The relation of the present approach to the traditional adiabatic
approach.

-The relation of the present solution to the known plane wave exact
solution.

The simplicity of the approach as compared with the adiabatic one,
promises to be very useful in extending it to more complicated situations,
like the system with different masses and systems with more particles,
currently under research, or systems in three dimensions modeled by
ZRP potentials. In the last case, the method can be applied to a wide
kind of systems to obtain asymptotic solutions which can be matched
to solutions obtained with methods like the R-matrix one, simplifying
substantially the calculations.

\section*{Acknowledgments}

Oak Ridge National Laboratory, is managed by UT-Battelle, LLC under
contract number DE-AC05-00OR22725. Support by the National Science
Foundation under grant number PHY997206 is gratefully acknowledged.
AAT thanks to CONACyT, project 4877-E for partial support and gratefully
acknowledges the financial support from the National Science Foundation
under grant number PHY997206 together with the hospitality of the
University of Tennessee. SYL also thanks the University for making
possible a visit. One of us, G.G. thanks the support under the PICT
Nro. 0306249 of the ANPCyT.

\noindent \bigskip
\bigskip
\textbf{APPENDIX A: Pseudo-Sturmian Functions} \bigskip

Fixing $\nu$, the general solution for the eigenvalue equation

\begin{equation}
\left[\frac{\partial^{2}}{\partial\theta^{2}}+\nu^{2}\right]\varphi(\theta)=\left[\rho\left(\nu\right)\frac{\pi c}{3}\sum_{j=0}^{5}\delta\left(\theta-\theta_{j}\right)\right]\varphi(\theta),\label{b1}\end{equation}
 with $\theta_{j}=\left(2j+1\right)\pi/6$, can be written as the
free angular wave solution $\varphi(\theta)=D_{\nu}\cos\left[\nu\left(\theta-\gamma_{j}\right)\right],\; j=0,1,...,5,$
provided that it be continuous through the boundary lines $\theta=\theta_{j}$
, that is, \begin{equation}
\cos\left[\nu\left(\theta_{j}-\gamma_{j+1}\right)\right]=\cos\left[\nu\left(\theta_{j}-\gamma_{j}\right)\right],\end{equation}
 and satisfies the boundary conditions

\begin{eqnarray}
\lim_{\zeta\rightarrow0} &  & \int_{\theta_{j}-\zeta}^{\theta_{j}+\zeta}\left\{ d\theta\left[\frac{\partial^{2}}{\partial\theta^{2}}+\nu^{2}\right]\varphi(\theta)\right.\nonumber \\
 &  & \left.\left.-\left[\rho\left(\nu\right)\frac{\pi c}{3}\sum_{l=0}^{5}\delta\left(\theta-\theta_{l}\right)\right]\varphi(\theta)\right\} =0,\right.\;\;\label{co1}\end{eqnarray}
 with $j=0,1,...,5~$. $D_{\nu}$, which does not depend on $j$ for
the symmetric solution, determined by normalizing the wave function\cite{Larsen}.
The requirement of continuity leads to the conditions $\gamma_{j}+\gamma_{j+1}=2\theta_{j}=\left(j+\left[j+1\right]\right)\pi/3$
or to $\gamma_{j-1}=\gamma_{j}$. The second condition does not satisfy
(\ref{co1}) so we shall use the first one, which can be written as
$\gamma_{j}=j\;\pi/3$. Now to focus on Eq.( \ref{co1}). Continuity
implies that the integral of the second term gives zero. For each
$j$, the first and the third terms give

\begin{eqnarray}
\lim_{\zeta\rightarrow0} &  & \left(\frac{\partial}{\partial\theta}\left(\cos\nu\left[\theta-\gamma_{j+1}\right]\right)_{\theta=\theta_{j}+\varsigma}\right.\nonumber \\
 &  & -\left.\frac{\partial}{\partial\theta}\left(\cos\nu\left[\theta-\gamma_{j}\right]\right)_{\theta=\theta_{j}-\varsigma}\right)\nonumber \\
 &  & -\rho\left(\nu\right)\frac{\pi c}{3}\cos\nu\left[\theta_{j}-\gamma_{j}\right]=0.\label{b4}\end{eqnarray}
 If we select a symmetric solution, then

\begin{equation}
-\frac{\partial}{\partial\theta}\cos\left(\nu\left[\theta_{j}+\varsigma-\gamma_{j+1}\right]\right)=\frac{\partial}{\partial\theta}\cos\left(\nu\left[\theta_{j}-\varsigma-\gamma_{j}\right]\right),\end{equation}
 and taking the limit in Eq.(\ref{b4}), we obtain the desired form
of the boundary condition

\begin{equation}
\lim_{\theta^{-}\rightarrow\theta_{j}}\frac{1}{\rho\left(\nu\right)\cos\left(\nu\left[\theta-\gamma_{j}\right]\right)}\frac{\partial}{\partial\theta}\cos\left(\nu\left[\theta-\gamma_{j}\right]\right)=-\frac{\pi c}{6},\label{bc1}\end{equation}
 where $\; j=0,1,...,5$. Calculating the derivative and the limit
in Eq. (\ref{bc1}) yields

\begin{equation}
\frac{6}{\pi c}v\tan\nu\pi/6=\rho\left(\nu\right).\label{r}\end{equation}

We conclude that $\cos\left[\nu\left(\theta-j\pi/3\right)\right],\; j=0,1,...,5$,
satisfies Eq.(\ref{b1}), provided that $\rho\left(\nu\right)$ satisfies
Eq.(\ref{r}).

\bigskip
{}\textbf{APPENDIX B: Derivation of the plane wave representation in
terms of cartesian coordiantes} \bigskip

For the $2+1$ system, and aside from an ultimate normalization, the
incoming wave function from Eq.(\ref{symf}) can be written in terms
of cartesian coordinates, as

\begin{eqnarray}
\psi^{i} & = & \exp\left\{ \frac{\pi c}{6}\left(\frac{x_{1}-x_{2}}{\sqrt{2}}\cos\left[\left(j-1\right)\frac{\pi}{3}\right]\right.\right.\nonumber \\
 &  & +\left.\frac{x_{1}+x_{2}-2x_{3}}{\sqrt{6}}\sin\left[\left(j-1\right)\frac{\pi}{3}\right]\right)\nonumber \\
 &  & -ik\left(\frac{x_{1}+x_{2}-2x_{3}}{\sqrt{6}}\cos\left[\left(j-1\right)\frac{\pi}{3}\right]\right.\nonumber \\
 &  & -\left.\left.\frac{x_{1}-x_{2}}{\sqrt{2}}\sin\left[-\left(j-1\right)\frac{\pi}{3}\right]\right)\right\} \nonumber \\
 &  & +\exp\left\{ -\frac{\pi c}{6}\left(\frac{x_{1}-x_{2}}{\sqrt{2}}\cos\left[\left(j+1\right)\frac{\pi}{3}\right]\right.\right.\nonumber \\
 &  & +\left.\frac{x_{1}+x_{2}-2x_{3}}{\sqrt{6}}\sin\left[\left(j+1\right)\frac{\pi}{3}\right]\right)\nonumber \\
 &  & -ik\left(\frac{x_{1}+x_{2}-2x_{3}}{\sqrt{6}}\cos\left[\left(j+1\right)\frac{\pi}{3}\right]\right.\nonumber \\
 &  & -\left.\left.\frac{x_{1}-x_{2}}{\sqrt{2}}\sin\left[-\left(j+1\right)\frac{\pi}{3}\right]\right)\right\} \end{eqnarray}
 Evaluating the trigonometric functions for $j=0$ in the above expression,
the argument of the first exponential function takes the form

\begin{eqnarray}
i\left[\right. & - & \frac{2}{\sqrt{6}}k\; x_{1}+\left(-i\frac{\pi c}{6\sqrt{2}}+i\frac{1}{\sqrt{6}}k\right)\; x_{2}\nonumber \\
 & + & \left.\left(i\frac{\pi c}{6\sqrt{2}}+i\frac{1}{\sqrt{6}}k\right)\; x_{3}.\right]\end{eqnarray}
 By labeling the particle wave numbers as in \cite{Amaya},

\begin{eqnarray}
k_{1} & = & i\frac{\pi c}{6\sqrt{2}}-\frac{1}{\sqrt{6}}k,\nonumber \\
k_{2} & = & -i\frac{\pi c}{6\sqrt{2}}-\frac{1}{\sqrt{6}}k,\nonumber \\
k_{3} & = & \sqrt{\frac{2}{3}}k,\end{eqnarray}

\noindent the incoming wave takes the form \begin{eqnarray*}
\psi^{i} & = & \exp\left\{ -i\left(k_{3}x_{1}+k_{2}x_{2}+k_{1}x_{3}\right)\right\} _{j=0}\\
 & + & \exp\left\{ i\left(k_{2}x_{1}+k_{3}x_{2}+k_{1}x_{3}\right)\right\} _{j=0}\end{eqnarray*}
 The outgoing wave for $j=0$ can be obtained from the incoming one
by substituting $k$ by $-k$, which in turns means interchanging
$k_{1}\leftrightharpoons k_{2}$ and inverting the sign of the whole
argument within all exponentials, that is,

\noindent \begin{eqnarray*}
\mathcal{S} & \left[\exp\left\{ -i\left(k_{1}x_{1}+k_{3}x_{2}+k_{2}x_{3}\right)\right\} _{j=0}\right.\\
+ & \left.\exp\left\{ i\left(k_{3}x_{1}+k_{1}x_{2}+k_{2}x_{3}\right)\right\} _{j=0}\right].\end{eqnarray*}
 The wave associated to the factor $\mathcal{S}_{3}$ can be written
as

\begin{center}$\begin{array}{cc}
exp & \left\{ -i\left(k_{1}x_{1}+k_{2}x_{2}+k_{3}x_{3}\right)\right\} _{j=0}\\
+ & \left.\exp\left\{ i\left(k_{2}x_{1}+k_{1}x_{2}+k_{3}x_{3}\right)\right\} _{j=0}\right].\end{array}$\end{center}

\noindent The above results correspond to the sector $j=0$ in the
$(\rho,\theta)$ plane, in which the order of particles is given by
$x_{2}<x_{3}<x_{1}$. The waves in different sectors can be obtained
by the appropriate permutation of the set of coordinates $\left\{ x_{1},x_{2},x_{3}\right\} $.
The completely symmetric wave plane may then be written as

\noindent \begin{eqnarray}
\psi=\Sigma_{p} & \left[\left\{ \exp\left[-i\left(k_{3}x_{1}+k_{2}x_{2}+k_{1}x_{3}\right)\right]_{j=j_{p}}\right.\right.\nonumber \\
 & +\left.\exp\left[i\left(k_{2}x_{1}+k_{3}x_{2}+k_{1}x_{3}\right)\right]_{j=j_{p}}\right\} \nonumber \\
+\mathcal{S}(p) & \left\{ \exp\left[-i\left(k_{1}x_{1}+k_{3}x_{2}+k_{2}x_{3}\right)\right]_{j=j_{p}}\right.\nonumber \\
 & \left.+\exp\left[i\left(k_{3}x_{1}+k_{1}x_{2}+k_{2}x_{3}\right)\right]_{j=j_{p}}\right\} \nonumber \\
+\mathcal{S}_{3}(p) & \left\{ \exp\left[-i\left(k_{1}x_{1}+k_{2}x_{2}+k_{3}x_{3}\right)\right]_{j=j_{p}}\right.\nonumber \\
 & +\left.\left.\exp\left[i\left(k_{2}x_{1}+k_{1}x_{2}+k_{3}x_{3}\right)\right]_{j=j_{p}}\right\} \right]\end{eqnarray}
 where the sum runs over all permutations of the set $\left\{ x_{1},x_{2},x_{3}\right\} $.


\begin{thebibliography}{10}
\bibitem{Evora}\textit{XVIIth European Conference on Few-Body Problems in Physics}.
Conference Handbook. Ed.: A. Stadler et al. Evora 2000.
\bibitem{NielsenMacek}E. Nielsen and J. H. Macek, Phys. Rev. Lett \textbf{83}, 1566 (1999).
\bibitem{Inouye}S. Inouye, M. R. Andrews, K Stenger, H.J. Miesner, D. M. Stamper-Kurn
and W. Ketterle, Nature (London) \textbf{392}, 151 (1998).
\bibitem{Moerdijk}A. J. Moerdijk, H. M. J. M. Boesten and B. J. Verhaar, Phys. Rev.
A 53, R916 (1996).
\bibitem{Esry}B. D. Esry, C. H.. Greene, Y. Zhou and C. D. Lin, J. Phys. B 29, L51
(1996).
\bibitem{Fedichev}P. O. Fedichev, M. W. Reynolds and G. V. Shlyapnikov, Phys. Rev. Lett.
77, 2921 (1996).
\bibitem{DemkovOstrovskii}Yu. N. Demkov and V. N. Ostrovsky, \textit{Zero-Range Potentials and
Their Applications in Atomic Physics}. Plenum Press, NY, 1988.
\bibitem{GasaneoMacek}G. Gasaneo and J. H Macek, Jour. Phys. B \textbf{35} 2239-2250 (2002).
\bibitem{GasaneoPRA}G. Gasaneo, S. Yu. Ovchinnikov and J. H. Macek, Phys. Rev. A submitted
(2002).
\bibitem{McG}J. B. McGuire, J. Math. Phys. \textbf{5}, (1964) 622.
\bibitem{Yang}C. N. Yang, Phys. Rev. Lett. \textbf{19}, 1312 (1967); C. N. Yang,
Phys. Rev. \textbf{168}, 1920 (1968).
\bibitem{Bethe}H. A. Bethe, Z. Phys. \textbf{71}, (1931) 205.
\bibitem{numerous}L. R. Dodd, J. Math. Phys. \textbf{11}, (1970) 207; H. B. Thacker,
Phys. Rev. D \textbf{11}, (1975) 838; A. Amaya-Tapia, S.Y. Larsen
and J. Popiel, Few-Body Sys. \textbf{23}, (1997) 87; S. I. Vinitsky,
S. Y. Larsen, D. V. Pavlov, D. V. Proskurin. Phys. At. Nucl. \textbf{64},
27 (2001); O. Chuluunbaatar, A. A. Gusev, S. Y. Larsen, S. I. Vinitsky.
J. Phys. \textbf{A35}, L513 (2002)
\bibitem{thomas}L. H. Thomas, Phys. Rev. \textbf{47}, 903 (1935). 
\bibitem{GOM}G. Gasaneo, S. Yu. Ovchinnikov and J. H. Macek, Jour. of Phys. \textbf{A34},
(2001) 8941.
\bibitem{Gradshteyn}I. S. Gradshteyn and I. M. Ryzhik, \textit{Table of Integrals, Series
and Products.} Academic Press, NY, 1965, p.773.
\bibitem{Amaya}A. Amaya-Tapia, S.Y. Larsen and J. Popiel, Few-Body Sys. \textbf{23},
(1997) 87; W. Gibson, S. Y. Larsen and J. J. Popiel, Phys. Rev. A
\textbf{35} (1987) 4919.
\bibitem{Larsen}S.Y. Larsen, in \textit{Few Body Methods: Principles and Applications}.
T. K. Lim, C. G. Bao, D. P. Hou and H. S. Huber, eds. p. 467, Singapore,
World-Scientific 1986.\end{thebibliography}
\end{document}